\documentclass{aa} 
\bibpunct{(}{)}{;}{a}{}{,}

\usepackage{graphicx}
\usepackage{txfonts}
\begin{document}
%
   \title{Finding halo streams with a pencil-beam survey:}

   \subtitle{new wraps in the Sagittarius stream}

   \author{B. Pila-D\'iez \inst{1}
	  \and
	  K. Kuijken \inst{1} 
	  \and
	  J.T.A. de Jong \inst{1}
          \and
	  H. Hoekstra \inst{1}
          \and
	  R.F.J. van der Burg \inst{1}}

   \institute{Leiden Observatory, Leiden University,
              Oort Building, Niels Bohrweg 2, NL-2333 CA Leiden\\
              \email{piladiez@strw.leidenuniv.nl}  \email{kuijken@strw.leidenuniv.nl}
	      \email{jelte@strw.leidenuniv.nl} \email{hoekstra@strw.leidenuniv.nl}
              \email{vdburg@strw.leidenuniv.nl}
             }

   \date{Received Month 00, 2013; accepted Month 00, Year}

 
  \abstract
   {We use data from two CFHT-MegaCam photometric pencil-beam surveys in the g' and 
    the r' bands to measure distances to the Sagittarius, the Palomar 5 and the 
    Orphan stream. We show that, using a cross-correlation algorithm to detect the 
    turnoff point of the main sequence, it is possible to overcome the main limitation 
    of a two-bands pencil-beam survey, namely the lack of adjacent control-fields that 
    can be used to subtract the foreground and background stars to enhance the signal on 
    the colour-magnitude diagrams (CMDs). We describe the cross-correlation algorithm 
    and its implementation. We combine the resulting 
    main sequence turnoff points with theoretical isochrones to derive photometric 
    distances to the streams. Our results (31 detections on the Sagittarius stream and 
    one each for the Palomar 5 and the Orphan streams) confirm the findings by previous 
    studies, expand the distance trend for the Sagittarius faint southern branch and, 
    for the first time, trace the Sagittarius faint branch of the northern-leading 
    arm out to 56 kpc.
    In addition, they show evidence for new substructure: we argue that these detections 
    trace the continuation of the Sagittarius northern-leading arm into the southern 
    hemisphere, and find a nearby branch of the Sagittarius trailing wrap in the northern 
    hemisphere.
    }  


   \maketitle
%
%
%

\section{Introduction}\label{intro}

   In the past decade our picture of the Milky Way's stellar halo has dramatically 
   changed thanks to the advent of several observational surveys, which have 
   shown the richness and complexity of the substructure in the Galactic halo 
   \citep{ibat01,newb02,maj03,yan03,martin04CMa,grilldion07b,belok06,belok07orphan, 
   belok07quintet}. Our Galaxy is still undergoing an assembling process, where part 
   of the infalling material has already been accreted and become dynamically relaxed 
   \citep{helmi99,sheff12}, part of it is still dynamically cold \citep{bell08,juric08} 
   and another part is in the process of being dynamically stripped or even 
   approaching its first dynamical encounter with our Galaxy \citep{kall06feb,
   kall06dec,piatek08,besla10,rocha12}. 
   
   The most prominent example of a currently ongoing disruption is that of the 
   Sagittarius stream (Sgr stream). Since its discovery in 1996 \citep{mat96}, 
   the stream has been mapped wrapping over $\pi$ radians 
   on the sky, first through 2MASS \citep{maj03} and later through 
   SDSS \citep{belok06,kopos12}. There is general agreement that it is the stellar 
   debris of a disrupting satellite galaxy, the Sagittarius dwarf galaxy \citep{ibat94}, 
   which is currently being accreted by the Milky Way \citep{velaz95,ibat97,no10}. 
   The stream is composed of the leading and the trailing tails of this disruption 
   event \citep{mat96,ibat01,dp01,md01,md04,maj03,belok06, belok13}, which wrap at least 
   once around the Galaxy but have been predicted to wrap more than once 
   \citep{penarrub10,law10}. In addition, a bifurcation and what resembles an extra 
   branch parallel to the main component of the Sgr stream have been discovered both 
   in the northern hemisphere \citep{belok06} and in the southern hemisphere 
   \citep{kopos12}. The origin of this bifurcation and the meaning of the two 
   branches are still debated: they could represent wraps of different age 
   \citep{fellh06}, they could have arisen due to the internal dynamics of the progenitor 
   \citep{penarrub10,penarrub11} or they could indeed be due to different 
   progenitors and a multiple accretion event \citep{kopos12}. 

   On the other hand, one of the simplest and neatest examples of a disrupting 
   satellite is that of the Palomar 5 globular cluster \citep{sand77,oden02,dehnen04} 
   and its stream \citep{oden01,oden03}. This stream extends over $20^{\mathrm{o}}$ 
   along its narrow leading and trailing tails. It displays an inhomogeneous stellar 
   density in what resembles gaps or underdensities \citep{grilldion06pal5}; 
   the origin of this stellar distribution has been attributed both to interactions 
   with dark satellites \citep{carlb12} and to epicyclic motions of stars along the 
   tails \citep{mb12}. 

   Finally, there are also cases of streams with unknown progenitors, such as the 
   so-called Orphan stream \citep{grill06orphan,belok06,belok07orphan,newb10orphan}. 
   This stream extends for $~50^{\mathrm{o}}$ in the North galactic cap, and the 
   chemical signatures from recent spectroscopic observations associate its 
   progenitor with a dwarf galaxy \citep{casey13I,casey13II}. A number of plausible 
   progenitors have been suggested \citep{zucker06,fellh07,jin07,sales08}, but it 
   is still possible that the true progenitor remains undiscovered in the southern 
   hemisphere \citep{casey13I}.
   
   In general, the discovery of most of the substructures in the halo of the 
   Milky Way has been possible thanks to photometric multi-colour wide area surveys. 
   Such surveys pose several advantages for this kind of search. First, their 
   multiple-band photometry allows for stellar population selections (halo or 
   thick disk; red clump, main sequence turnoff point, etc.) based on colour-colour 
   stellar loci. These selection criteria can be used to make stellar density maps 
   that track the streams all through the survey's coverage area \citep{maj03,
   belok06}. Second, their continuous coverage of a large area allow the fields 
   adjacent to the substructure to act as control fields. In this way, the 
   colour-magnitude diagrams (CMDs) of the control fields can be used to statistically 
   subtract the foreground and the background stars from the fields probing the 
   substructure. This enhances the signature of the stellar population belonging 
   to the stream or satellite (by removing the noise), and makes it possible to 
   identify age and distance indicators such as the red clump or the main sequence 
   turnoff point \citep{belok06,kopos12,slater13}.
   
   In this paper we explore the possibilities of using deep two-band 
   pencil-beam surveys instead of the usual wide-area multi-colour surveys in order 
   to detect and characterize stellar streams of the halo and, in particular, we revisit 
   the Sagittarius, the Palomar 5 and the Orphan streams. We derive photometric distances 
   using purely the main sequence turnoff point and --unlike other works-- regardless of 
   the giant branch and its red clump.

%
%

\section{Observations and data processing }\label{data}


   \subsection{Description of data set}\label{subsec:dataset}
	
   We use deep photometric imaging from the MENeaCS and the CCCP surveys 
   \citep{sand12meneacs,hoekstra12cccp,bildfell12combined} as well as several additional 
   archival cluster fields, observed with the CFHT-MegaCam instrument. These surveys 
   targeted pre-selected samples of galaxy clusters; therefore the surveys geometry takes 
   the form of a beam-like survey where the pointings are distributed without prior 
   knowledge of the halo substructure (blind survey). 

   \begin{figure*}
   \centering
   \includegraphics[width=\textwidth]{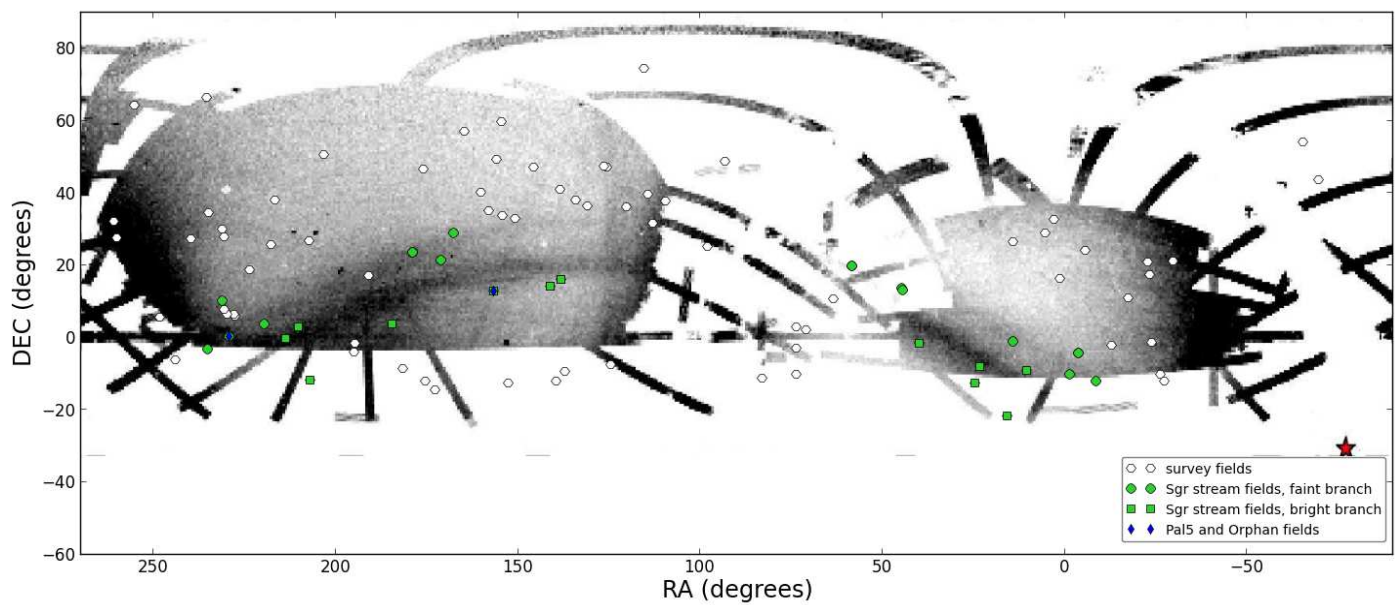}
      \caption{Equatorial map showing the position of all the fields from our survey (white 
               hexagons) and highlighting the ones that lay on the Sagittarius stream 
               (green circles for the faint branch and green squares for the bright branch), 
               on the Palomar 5 stream and on the Orphan stream (blue diamonds).
               The background image is the SDSS-DR8 map of the Sgr stream from \citet{kopos12}.
               }
         \label{fig:mapPointings}
   \end{figure*}

   Our pointings are one square degree wide and spread over the sky visible to CFHT. Each 
   consists of several exposures through the g' and r' filters with image quality of of 
   sub-arcsecond seeing. After stacking the individual exposures, the limiting magnitudes 
   reach $\sim25.0$ at the $5.0\sigma$ level. Out of 
   the 97 fields, at least 25 fall on the structure of the Sagittarius (Sgr) stream 
   and show distinct signatures in their CMDs, 
   one on the Orphan stream, 
   one on the Palomar 5 stream and 
   three to seven on the Virgo Overdensity and the Virgo Stellar Stream \citep{duffau06,
   juric08,casetti09,prior09,bonaca12virgo} (see figure~\ref{fig:mapPointings}). 
   Further away from the plane of the Sgr stream, we also find 
   three fields to be coincident with the Triangulum-Andromeda structure 
   \citep{rochapinto04,bonaca12triangulum}, 
   two to three with the Pisces Overdensity \citep{watkins09,sesar10,sharma10}, 
   one transitional between the Triangulum-Andromeda and the Pisces Overdensity, 
   four with the Anticenter Structure \citep{grillmair06anticenter} and 
   two to three with the NGC5466 stream \citep{grillmairjohnson06,fellhauer07ngc5466}.  
   We also find 
   two fields on the Lethe stream \citep{grillmair09new4treams}, 
   four on the Styx stream \citep{grillmair09new4treams}, 
   one on a region apparently common to the Styx and Cocytos streams 
   \citep{grillmair09new4treams} and 
   two on the Canis Major overdensity \citep{martin04CMa}.

   In this paper we concentrate on the clearest structures (those where the 
   contrast-to-noise in the CMD is higher) in order to test the capabilities of our 
   method. In particular, we address the Sagittarius stream, the Palomar5 stream and 
   the Orphan stream.


   \subsection{Correction of the PSF distortion and implications for 
      the star/galaxy separation}\label{subsec:PSF}
 
   Before building catalogues and in order to perform an accurate star/galaxy separation, 
   it is necessary to rectify the varying PSF across the fields of the CFHT images. 

   In order to correct for this effect, we make use of a 'PSF-homogenizing' code 
   (K. Kuijken et al., in prep.). The code uses the shapes of bright objects 
   unambiguously classified as stars to map the PSF across the image, and then convolves it 
   with an appropriate spatially variable kernel designed to render the PSF gaussian 
   everywhere. 
   With a view to obtaining a PSF as homogeneous as possible, we treat the data as follows 
   \citep{vdBurg13}: 
   i) we implement an accurate selection of sufficiently bright stars based on an 
   initial catalogue, 
   ii) we run the code on the individual exposures for each field, and 
   iii) we reject bad exposures based on a seeing criterion \footnotemark[1] before stacking 
   them into one final image, on which we perform the final source extraction and photometry.
   \footnotetext[1]{
   The rejection of exposures derives from trying to optimize the image quality while 
   achieving the desired photometric depth. Thus our seeing criterion is a variable number 
   dependent on the field itself, the seeing distribution for individual exposures and the 
   individual plus total exposure time. In general it takes values $<\approx0.9$.}

\begin{figure}
   \centering
   \includegraphics[width=95mm]{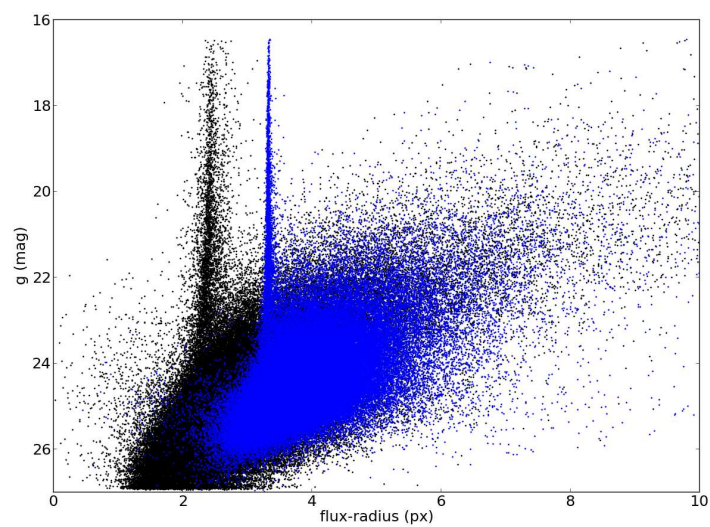}
      \caption{Brightness versus size diagram of all the sources in one of our pointings. 
               The stellar locus prior to the PSF-homogenization (black) is wider and 
               therefore subject to greater galaxy contamination at the faint end than the 
               stellar locus posterior to the correction (blue) because the PSF initially 
               varies across the field.}
         \label{fig:GaussedSize}
   \end{figure}

\begin{figure}
   \centering
   \includegraphics[width=95mm]{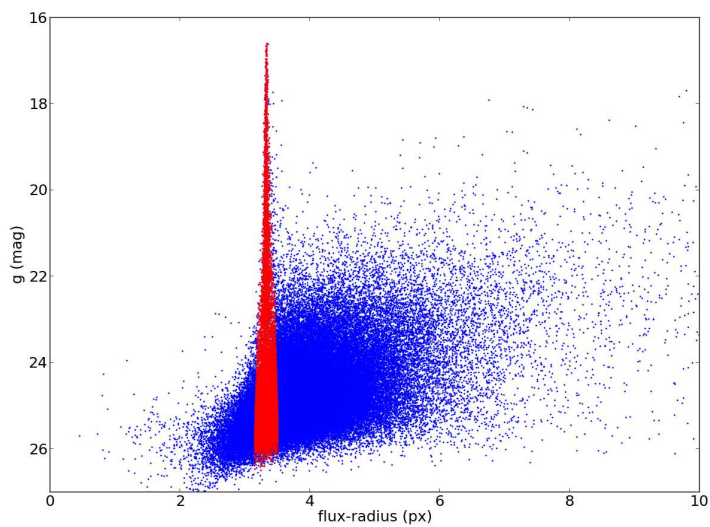}
      \caption{Brightness versus size diagram showing all the PSF-corrected sources 
               (blue) and the subset of sources selected as stars through our star/galaxy 
               separation algorithm (red) for one of our pointings. Although the star 
               selection may not be complete at the faint end due to increasing scatter, 
               our algorithm minimizes the galaxy contamination, which otherwise would be 
               the main obstacle for detecting faint structures in the CMD.}
         \label{fig:starsSelec}
   \end{figure}

   The advantages of this procedure are twofold. First, because the resulting PSF for each 
   exposure is gaussian, all the stars become round.
   Second,  because the PSF anisotropy is removed from all exposures before stacking, the 
   dispersion in size for the point-source objects becomes smaller, even if the average 
   value increases after stacking the individual exposures (see figure~\ref{fig:GaussedSize}).  
   These two improvements significantly reduce the galaxy contamination when performing 
   the star selection (illustrated in figure~\ref{fig:starsSelec}).
   Additionally, homogenizing the PSF also allows to measure colours in fixed apertures.

   From the final images, we extract the sources and produce photometric catalogues 
   using SExtractor \citep{bertinSExtractor}.
   To derive the stellar catalogues, we use a code that filters the source catalogues 
   as follows:
   i) finds the saturated stars and removes them from the stellar catalogue; 
   ii) evaluates the distribution of bright sources ($r'=[18.0,20.0]\ \mathrm{mag}$) in 
   the brightness-size parameter space, assumes a gaussian distribution in the size and 
   in the ellipticity parameters ($e_1$, $e_2$)\footnotemark[2] of stars, and uses this 
   information to define the boundaries of the stellar locus along the bright range;
   iii) evaluates the dependence of the width of the stellar locus on brightness and 
   extrapolates the relation to fainter magnitudes; 
   iv) applies the extended stellar locus and an ellipticity criterion to drop galaxies 
   from the stellar catalogue.
   
   \footnotetext[2]{
      \begin{displaymath}
         e_1 = \frac{1-q^2}{1+q^2}\cos{2\theta}    \, , \; \;   
         e_2 = \frac{1-q^2}{1+q^2}\sin{2\theta}    \, , \; \;  
      \end{displaymath}
      where $q=$axis ratio, $\theta=$position angle of the major axis.

   }

\begin{figure*}
   \centering
   \includegraphics[width=\textwidth]{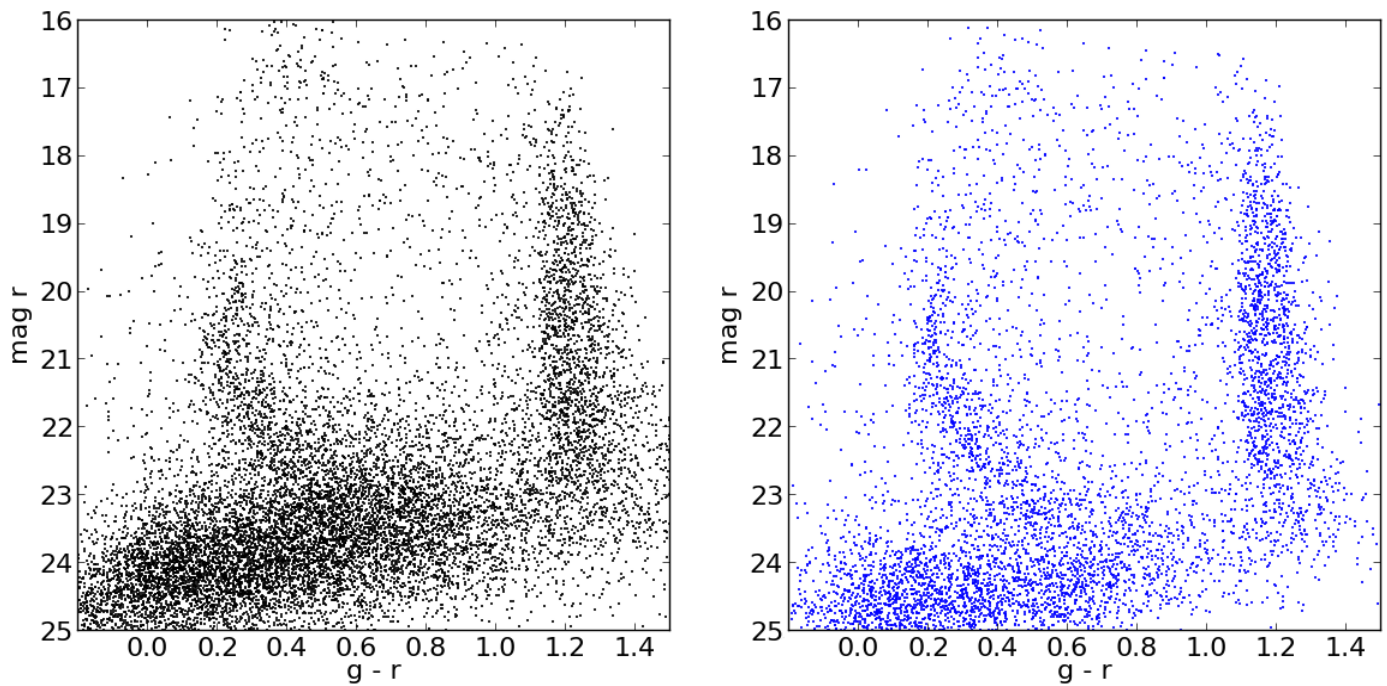}
      \caption{Colour-Magnitude Diagram (CMD) displaying the selection of sources 
               considered stars (selected as explained in section 2.2). The plume on 
               the red side ($g-r\approx1.2$) is composed of the nearby M-dwarfs, whereas 
               the main sequence on the bluer side ($0.18<g-r<0.6$) corresponds to a halo 
               overdensity located at a particular well-defined distance. 
               The cloud of sources at faint magnitudes are faint galaxies that enter the 
               star selection. 
               \emph{Left:} CMD derived from an image that has not been PSF-corrected. 
               \emph{Right:} CMD derived from a PSF-corrected image. After homogenizing 
               the PSF, the galaxy contamination decreases markedly below $r \approx 22$.}
         \label{fig:GaussedCMD}
   \end{figure*}

   For the stars resulting from this selection (figure~\ref{fig:starsSelec}), we correct 
   their photometry from galactic reddening by using the extinction maps from 
   \citet{schlegel98dustmaps}. The final stellar catalogues are used to build the CMDs 
   employed for our analysis. The PSF-corrected catalogues yield much cleaner CMDs than 
   the catalogues with similar star/galaxy separation but no PSF-correction 
   (figure~\ref{fig:GaussedCMD}).
   
   \subsection{Identification of the main sequence turnoff point}\label{subsec:turnoffpoint}
   
   The photometric depth of our data allows us to detect a number of halo substructures 
   several magnitudes below their main sequence turn-off point. However, because our 
   survey is a pencil-beam survey lacking control fields adjacent to our target fields, 
   we have no reference-CMDs representing a clean foreground plus a smooth halo, and thus 
   a simple foreground subtraction is not possible. Instead the halo substructures in our 
   survey can only be detected in those fields where the contrast in density between the main
   sequence stream stars and the foreground and background stars is significant in the CMD.

\begin{figure*}
   \centering
   \includegraphics[width=\textwidth]{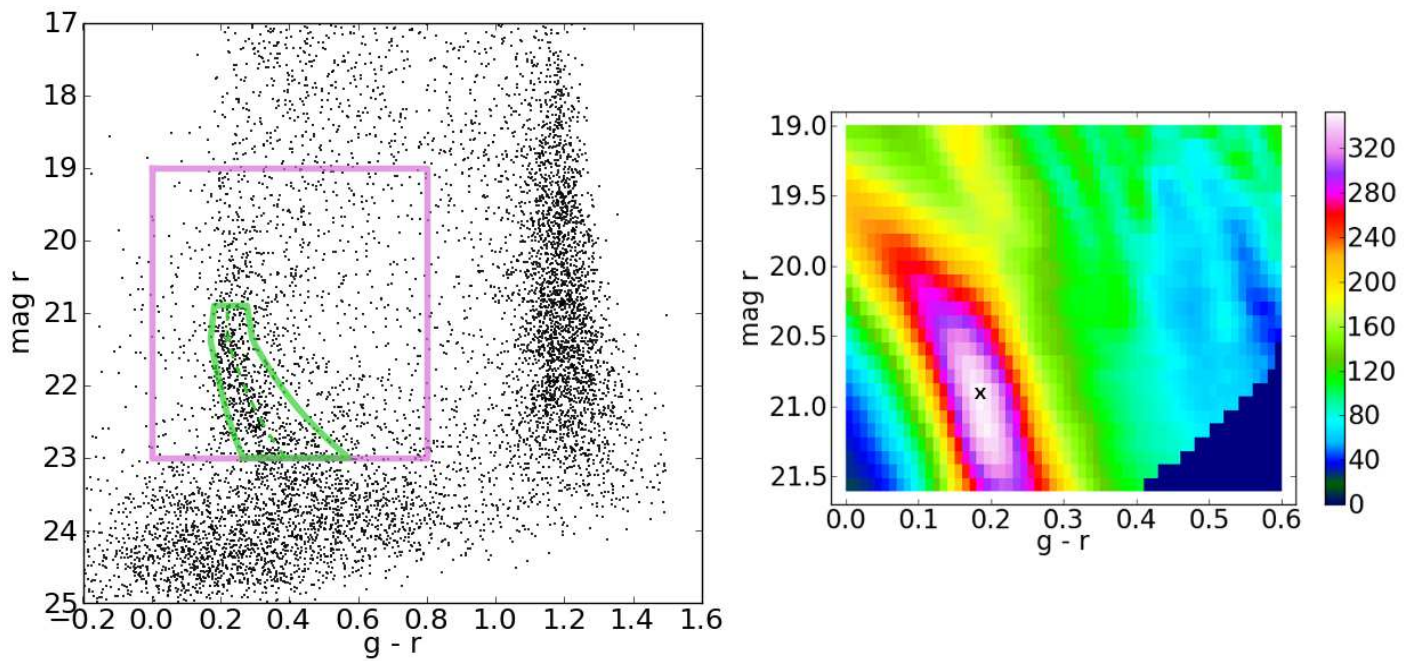}
      \caption{\emph{Left}: Dereddened CMD (black dots) with the search region (pink 
               solid-line rectangle) for the cross-correlation and the template main
               sequence-shaped function (green solid line) at the position of maximum 
               density (peak of the cross-correlation). 
               \emph{Right}: Binned diagram representing the weighted density of stars 
               resulting from the cross-correlation process. The density in each bin 
               corresponds to the integral of the template main sequence-shaped 
               function with top left corner in the position of the bin.}
         \label{fig:ccCMDbins}
   \end{figure*}

   Thus, in order to search for main sequences in the CMDs, we build a cross-correlation 
   algorithm that runs across a region of the CMD (the 'search region'), focused on the 
   colour range associated with the halo turnoff stars ($0.18 \leq g-r \leq 0.30$). Within 
   the boundaries of this search region, we slide a template main sequence-shaped 2D 
   function that operates over the number of stars and, for each step, yields an integral 
   representing the weighted density of stars in such a main sequence-shaped area. When the 
   template main sequence function coincides with a similarly shaped 
   overdensity in the CMD), the value of the cross-correlation (the weighted density) is 
   maximized, and a value for the turnoff point is assigned. This process is illustrated  
   in figure~\ref{fig:ccCMDbins}. 
   
   In some cases a CMD presents more than one main sequence signature with sufficient 
   contrast to noise. When this happens we use the detection of the primary 
   main sequence (the position of its turnoff point and its characteristic width-function) 
   to randomly subtract a percentage of the stars associated with it (lowering its 
   density to the foreground level) and detect the next prominent main sequence feature. 
   We name these main sequence detections as primary, secondary, etc., ranked by their 
   signal to noise. We require the signal to noise to be $>3.5\sigma$ for primary MSs and 
   $>4\sigma$ for the secondary or tertiary MSs after partially removing the primary one.

   \subsubsection{Shape of the template main sequence function}

   When constructing the template main sequence-shaped 2D function (from now on, 'template-MS'), 
   we use two ingredients. 
   The first one is a theoretical isochrone\footnotemark[3] of age $t = 10 \mathrm{Gyr}$ and 
   metallicity $[Fe/H] = -1.58$, which is used to define the central spine of the template-MS. 
   The position of this central spine is later shifted in magnitude and colour steps during 
   the cross-correlation. 
   Since we are only interested in the shape of this isochrone (its absolute values are 
   irrelevant because it will be shifted) and since we are searching for halo substructures, 
   we choose the above age and metallicity values because they yield an isochrone shape 
   representative of old metal-poor stellar populations. 
   The second ingredient is a magnitude-dependent colour-width, which is used to broaden the 
   isochrone template as illustrated in the left panel of fig.~\ref{fig:ccCMDbins}).
   
   \footnotetext[3]{Through all this work we use a subset of theoretical isochrones from 
   http://stev.oapd.inaf.it/cmd. The theoretical isochrones (\citet{marigo08}, with 
   the corrections from Case A in \citet{girardi10} and the bolometric corrections 
   for Carbon stars from \citet{loidl01}) are provided as observable quantities 
   transformed into the CFHT photometric system.}

   The width is in general directly derived from the width of the locus of nearby M-dwarfs 
   ($1.0<g-r<1.4$). The width of this feature is calculated for a number of magnitude bins 
   as three times the standard deviation in colour for each bin. Then a functional form 
   dependent on magnitude is obtained through polynomial fitting. In a few cases, minor 
   tweaking is needed to compensate for extremely large widths (colour shifts become 
   insensitive to any substructure) or for extremely small widths (density values become 
   meaningless due to the built-in weight [see below]). This way of defining the width of 
   the template-MS accounts for the observational broadening of intrinsically well defined 
   stellar loci due to increasing photometric uncertainties at faint magnitudes.

   \subsubsection{Weights within the template MS-function}

   In addition to a theoretically and observationally motivated shape for the template-MS, 
   we also give a different weight to each region of the template. This means that, for each 
   step of the cross-correlation, the stars contained within will contribute differently to 
   the enclosed stellar density depending on how far they are from the spine of the template-MS. 
   
   The weight in colour (stars near the spine of the template-MS are more likely to belong 
   to the main sequence than stars close to the boundaries) is assigned through the exponential 
   term in a gaussian weight function. 
   We match the standard deviation of the gaussian weight to the standard deviation of the 
   template-MS width ($3\sigma=\omega_{MS}$) so that all the stars contained within the 
   template-MS are assigned a weight. 
   To guarantee that the weight does not favour bright features, we choose the amplitude of the 
   gaussian function to be such that the integral of the weight function between the edges 
   of the template-MS function is the same for all magnitudes. 
   
   The resulting weight function for a given star in the template-MS at a particular 
   step of the cross-correlation then follows:
   \begin{equation}
         \ \ W_{\ast}(mag,colour) =\frac{A}{\sqrt{2\pi}\sigma(mag)}\cdot exp\left\{-\frac{[colour-\eta_{CC}(mag)]^2}{2[\sigma(mag)]^2}\right\}
   \end{equation}
   where $mag$ and $colour$ are the magnitude and colour of the weighted star, 
   $\eta_{CC}(mag)$ represents the theoretical isochrone at that particular step 
   of the cross-correlation, 
   and $\sigma(mag)=\frac{1}{3}\omega_{MS}(mag)$ is proportional to the width of the template-MS 
   function for that particular CMD.


   \subsection{Uncertainties in the turnoff point}\label{subsec:uncertainties}

   The colour and magnitude values for the turnoff point of a given main sequence, ($c_{TO}$, 
   $mag_{TO}$), are derived from the position of the template at which the cross-correlation 
   peaks. Therefore the uncertainties for these turnoff point values derive from the contribution 
   of individual stars to the position and shape of the main sequence (the uncertainty from 
   the CMD itself). 
   To evaluate this uncertainty, we carry out a bootstrapping process. In this process first 
   we generate re-sampled stellar catalogues by randomly withdrawing stars from one of our 
   true catalogues. Second we run the cross-correlation and obtain the turnoff points for each 
   of these re-samples. Third we consider the offsets between these turnoff points and the original 
   turnoff point and derive the standard deviation of the distribution. The contribution of any 
   CMD to the uncertainty of its turnoff point can then be calculated as a function of a 
   reference (bootstrapped) standard deviation, $s$:
   \begin{displaymath}
      \ \ \ \ E_{\mathrm{mag,CMD}} =\ f_{\mathrm{mag,BS}}\cdot\frac{(s_{{\mathrm{mag,BS}}})}{\left.\frac{\partial^2 \rho_{\mathrm{CC}}}{\partial^2 mag}\right|_{\mathrm{TO}}}    \, , \; \; \;  
      E_{\mathrm{c,CMD}} =\ f_{\mathrm{c,BS}}\cdot\frac{(s_{{\mathrm{c,BS}}})}{\left.\frac{\partial^2 \rho_{\mathrm{CC}}}{\partial^2 c}\right|_{\mathrm{TO}}} \, , \; \; 
   \end{displaymath}

   where, in practice, $s_{\mathrm{mag,BS}}$ and $s_{\mathrm{c,BS}}$ are the standard deviations 
   calculated for a number of representative fields, $f_{\mathrm{mag,BS}}$ and $f_{\mathrm{c,BS}}$ 
   are scale factors that allow to obtain the uncertainty for any field from the standard 
   deviation of the bootstrapped fields, and  $\frac{\partial^2 \rho_{\mathrm{CC}}}{\partial^2 mag}$ 
   and $\frac{\partial^2 \rho_{\mathrm{CC}}}{\partial^2 c}$ evaluate the prominence of the 
   particular overdensity as a function of magnitude or as a function of colour. In practice, 
   $E_{\mathrm{mag,CMD}}=s_{{\mathrm{mag,BS}}}$ and $E_{\mathrm{c,CMD}}=s_{{\mathrm{c,BS}}}$ for 
   the bootstrapped fields used as a reference.

   The photometric turnoff point distances are derived from the distance modulus. Therefore the 
   uncertainties in the distances can be calculated as a combination of two sources of error: 
   the uncertainty derived from the observed brightness of the turnoff point ($E_{\mathrm{mag,CMD}}$, 
   discussed above) and the uncertainty derived from the absolute brightness of the 
   turnoff point, which depends on the choice of isochrone (and thus on the uncertainty in the 
   age or in the metallicity).
   \begin{eqnarray}
      E_{\mathrm{\mu,TO}} & = & \sqrt{E_{\mathrm{mag,CMD}}^2 + E_{\mathrm{mag,isoch}}^2} \,;
   \end{eqnarray}

%
%

\section{The Sagittarius stream}\label{results}


   \subsection{Turnoff point distances to the Sgr stream}\label{subsec:resultsSgr}
   
   The Sagittarius stream is clearly probed by at least 25 of our 97 fields (see the red 
   and pink squares in figure~\ref{fig:mapPointings}). They probe both the faint and the 
   bright branches of the stream (the faint branch lying to the North of the bright one) 
   and also two transitional areas, indicating that the transversal drop in stellar 
   counts between both branches is not dramatic. 
   Some of these fields present more than one main sequence in their CMDs; for those fields 
   the secondary turnoff points are calculated by subtracting the primary MS and 
   re-running the cross-correlation (as explained in section~\ref{subsec:turnoffpoint}).
   
   Based on the turnoff point values obtained from the cross-correlation, we calculate 
   the distances to the Sagittarius stream in these 25 fields for 31 detections. For this 
   calculation, we assume a single stellar population represented by a theoretical 
   isochrone with age $t_{age} = 10.2 \ \mathrm{Gyr}$ and metallicity $[Fe/H]=-1.0 \ \mathrm{dex}$ 
   (for a detailed description on the set of isochrones see footnote~2 in 
   section~\ref{subsec:turnoffpoint}). We choose these age and metallicity values because 
   they match the age-metallicity relation for the Sgr dwarf galaxy \citep{layden00} 
   --which is also expected to hold for its debris-- and are consistent with the range that 
   characterizes old metal-poor populations. 
   
\begin{figure*}
   \centering
   \includegraphics[width=\textwidth]{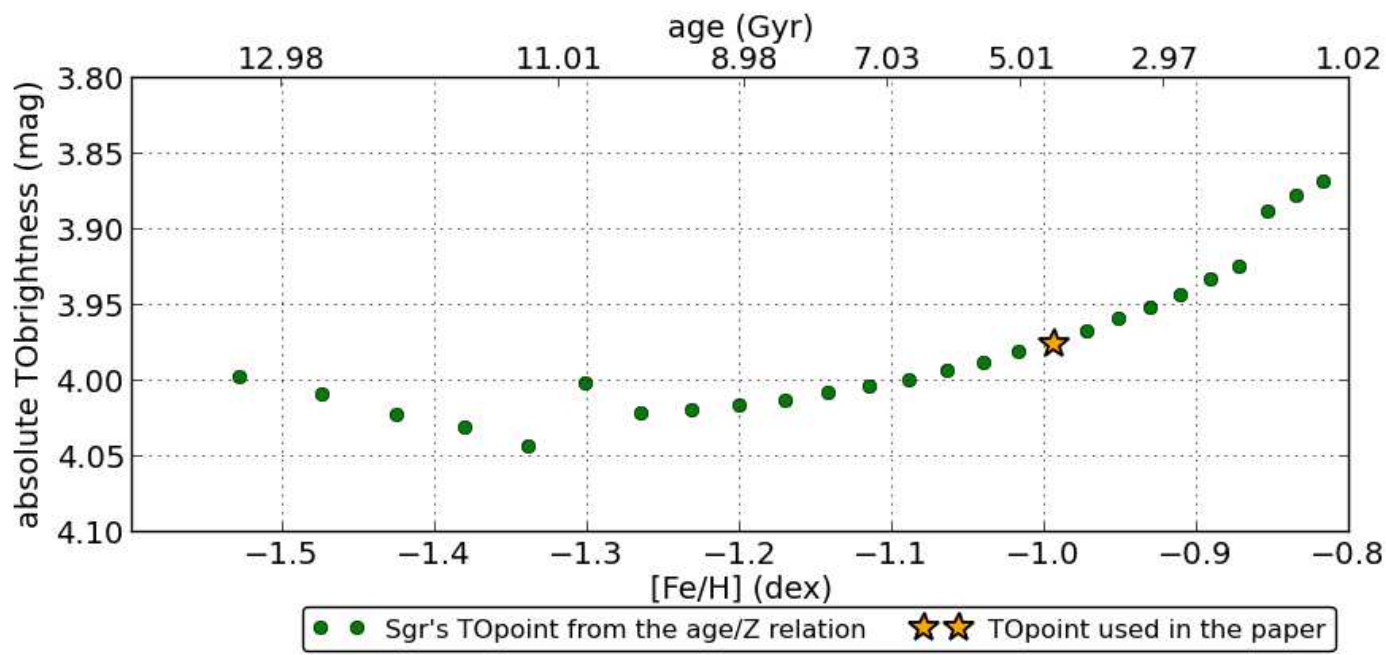}
      \caption{Absolute brightness of the turnoff point in the r band as a function of 
               metalllicity and age for metal poor populations (green circles). 
               The values in this diagram meet the age/metallicity relation for the Sgr 
               dwarf galaxy from \citet{layden00}. The isochrone used in this paper to 
               derive distances to the Sgr stream is represented with a yellow star,
               and its maximum difference to the other brightness values in this 
               range is $\Delta M=\pm0.1\mathrm{mag}$.}
       \label{fig:metallicitySgr}
   \end{figure*}

   To account for the potential influence on our distance measurements of a possible 
   metallicity gradient along the different Sgr arms \citep{chou07,shi12,vivas05,carlin12}, 
   we analyse the dependency of the 
   isochrones turnoff point absolute brightness ($M_{TO}$) with metallicity throughout 
   the Sgr metal-poor range (see figure~\ref{fig:metallicitySgr}). We find that for 
   $-1.53 \ \mathrm{dex} <[Fe/H]< -0.8 \ \mathrm{dex}$ the absolute brightness remains 
   nearly constant in the r band, with a maximum variation of $\Delta M=\pm0.1 \ \mathrm{mag}$. 
   We conclude that if we take this variation in absolute brightness as the isochrone 
   uncertainty in the distance modulus ($E_{\mathrm{mag,isoch}}=\Delta M$), we can use 
   the $t_{age} = 10.2 \ \mathrm{Gyr}$ and $[Fe/H]=-1.0 \ \mathrm{dex}$ isochrone to 
   calculate distances to any region of the Sgr stream. 

   \begin{table*}
   \caption{Position and distances to the Sgr stream,                      
            together with a tag for faint or bright branch membership, a tentative 
            classification as leading or trailing arm and a number specifying the hierarchy 
            of the detection in the CMD (primary, secondary,etc.). The distances are 
            indicated both as distance modulus and as heliocentric distance, with the 
            distance uncertainty ($E_{\mathrm{d}}$) in kpc. 
            }                      
   \label{table:distSgr}                          
   \centering                                                    
   \begin{tabular}{l r c c r c c c c}                        
   \hline\hline                                           
    Field & arm & detection & RA (deg) & DEC (deg) & $\mu (mag)$ & $d$ (kpc) & $E_{\mathrm{d}}$ (kpc) \\  
   \hline                                                  
      A2104$^{\mathrm{f}}$    & lead  & 1 & 235.040644 & -3.33158  & 18.8 & 56.6 & 3.1  \\ 
      RXJ1524$^{\mathrm{f}}$  & trail & 1 & 231.170583 &  9.93498  & 16.2 & 17.1 & 2.0  \\ 
      A2050$^{\mathrm{f}}$    & lead  & 1 & 229.080749 &  0.08773  & 18.7 & 54.1 & 8.7  \\ 
      A1942$^{\mathrm{f}}$    & lead  & 1 & 219.654126 &  3.63573  & 18.7 & 54.1 & 3.7  \\ 
      A1882$^{\mathrm{b}}$    & lead  & 1 & 213.667817 & -0.30598  & 18.5 & 49.3 & 5.7  \\ 
      A1835$^{\mathrm{b}}$    & lead  & 1 & 210.259355 &  2.83093  & 18.4 & 47.1 & 4.2  \\ 
      RXJ1347$^{\mathrm{b}}$  &   ?   & 1 & 206.889060 & -11.80299 & 15.5 & 12.4 & 7.3  \\ 
      ZwCl1215$^{\mathrm{b}}$ &   ?   & 1 & 184.433196 &  3.67469  & 16.7 & 21.5 & 2.9  \\ 
      ZwCl1215$^{\mathrm{b}}$ &   ?   & 3 & 184.433196 &  3.67469  & 15.0 & 9.8  & 2.6  \\ 
      A1413$^{\mathrm{f}}$    & lead  & 1 & 178.842420 &  23.42207 & 17.5 & 31.1 & 2.7  \\ 
      A1413$^{\mathrm{f}}$    & trail & 2 & 178.842420 &  23.42207 & 16.2 & 17.1 & 1.9  \\ 
      A1246$^{\mathrm{f}}$    & lead  & 1 & 170.987824 &  21.40913 & 17.6 & 32.6 & 9.2   \\ 
      A1185$^{\mathrm{f}}$    &   ?   & 1 & 167.694750 &  28.68127 & 16.3 & 18.7 & 12   \\ 
      ZwCl1023$^{\mathrm{b}}$ &   ?   & 1 & 156.489424 &  12.69030 & 17.4 & 29.7 & 11   \\ 
      A795$^{\mathrm{b}}$     & lead  & 1 & 141.030063 &  14.18190 & 16.0 & 14.2 & 2.8  \\ 
      A795$^{\mathrm{b}}$     &   ?   & 2 & 141.030063 &  14.18190 & 15.6 & 14.2 & 2.8  \\ 
      A763$^{\mathrm{b}}$     &   ?   & 1 & 138.150298 &  15.99992 & 16.7 & 21.5 & 2.6  \\ 
      A763$^{\mathrm{b}}$     & lead  & 2 & 138.150298 &  15.99992 & 15.8 & 14.2 & 1.0  \\ 
      RXJ0352$^{\mathrm{f}}$  & lead  & 1 & 58.263173  &  19.70387 & 15.7 & 13.6 & 0.7  \\ 
      RXJ0352$^{\mathrm{f}}$  & trail & 2 & 58.263173  &  19.70387 & 17.7 & 34.1 & 4.3  \\ 
      A401$^{\mathrm{f}}$     & trail & 1 & 44.759558  &  13.58975 & 17.4 & 29.7 & 3.4  \\ 
      A399$^{\mathrm{f}}$     & trail & 1 & 44.478652  &  13.05185 & 17.6 & 32.6 & 11   \\ 
      A370$^{\mathrm{b}}$     & trail & 1 & 39.963713  & -1.65806  & 17.6 & 32.6 & 4.8  \\ 
      A223$^{\mathrm{b}}$     & trail & 1 & 24.557005  & -12.77010 & 17.0 & 24.7 & 1.7  \\  
      RXJ0132$^{\mathrm{f}}$  & trail & 1 & 23.169048  & -8.04556  & 17.1 & 25.9 & 2.3  \\   
      A133$^{\mathrm{b}}$     & trail & 1 & 15.673483  & -21.88113 & 16.6 & 20.6 & 2.4  \\  
      A119$^{\mathrm{f}}$     & trail & 1 & 14.074919  & -1.23337  & 16.9 & 23.6 & 2.9  \\  
      A85$^{\mathrm{b}}$      & trail & 1 & 10.469662  & -9.28824  & 16.9 & 23.6 & 1.6  \\ 
      A2670$^{\mathrm{f}}$    & trail & 1 & 358.564313 & -10.40142 & 16.6 & 20.6 & 1.1  \\  
      RXJ2344$^{\mathrm{f}}$  & trail & 1 & 356.059633 & -4.36345  & 16.7 & 21.5 & 5.6  \\   
      RXJ2344$^{\mathrm{f}}$  & lead  & 2 & 356.059633 & -4.36345  & 15.6 & 13.0 & 1.2  \\   
      A2597$^{\mathrm{f}}$    & trail & 1 & 351.336736 & -12.11193 & 16.9 & 23.6 & 1.4  \\  
   \hline                                                             
   \end{tabular}
\begin{list}{}{}
\item[$^{\mathrm{b}}$] Bright branch
\item[$^{\mathrm{f}}$] Faint branch
\end{list}
   \end{table*}

\begin{figure*}
   \centering
   \includegraphics[width=\textwidth]{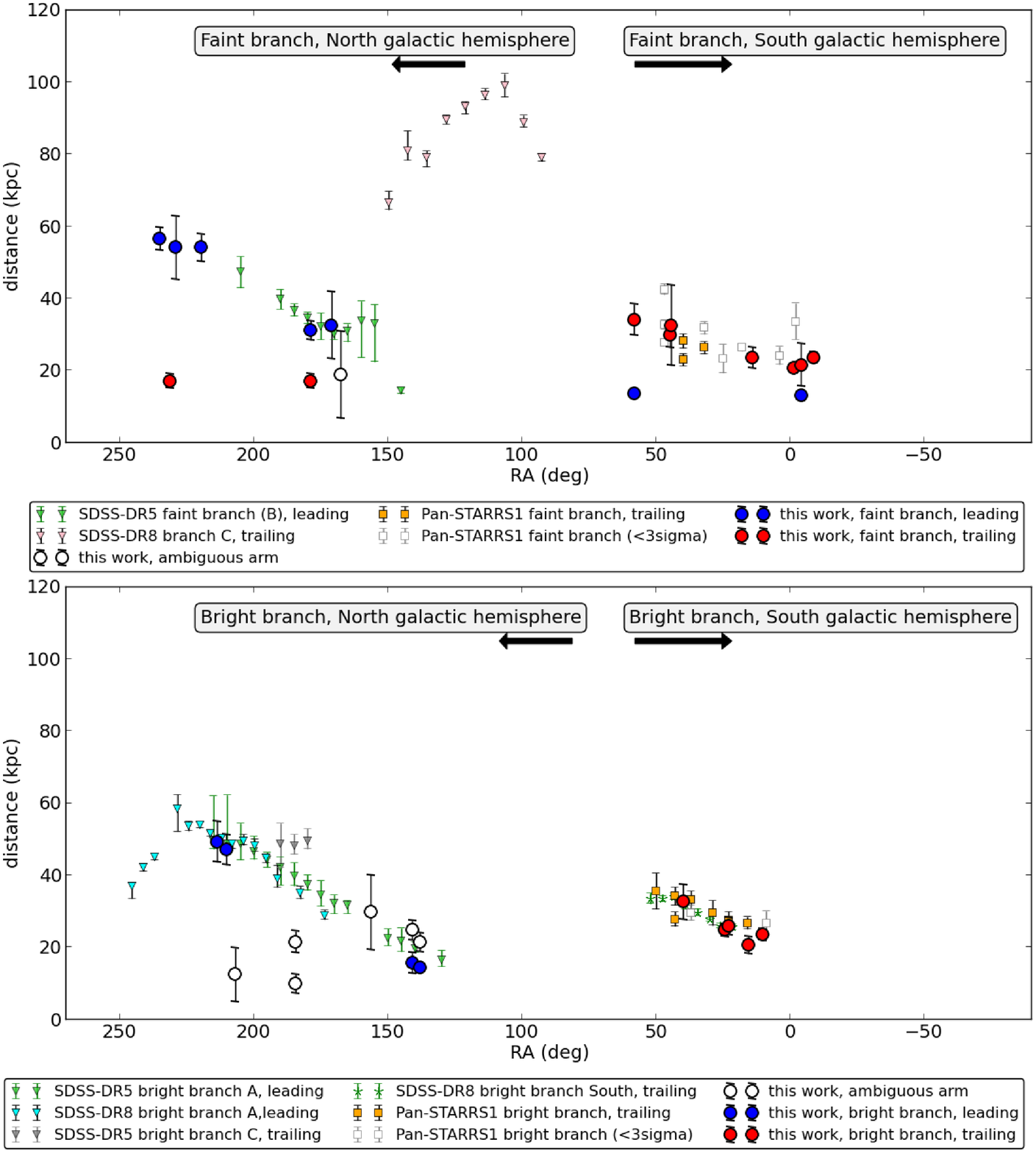}
      \caption{Photometric main sequence turnoff point distances for the Sagittarius 
               stream along right ascension (northern-leading tail and southern-
               trailing tail). The top panel shows results for the faint 
               branch, whereas the bottom panel corresponds to the bright arm. Our data 
               (blue circles for leading tails and red circles for trailing tails) are 
               based on the theoretical isochrones by 
               \citet{marigo08} and the corrections by \citet{girardi10}, for a 
               $10.2 \ \mathrm{Gyr}$ old stellar population with $[Fe/H]=-1.0$. 
               Other distance values correspond to \citet{belok06} (green and grey triangles), 
               \citet{kopos12,kopos13erratum} (green asterisks), \citet{belok13} (pink triangles) 
               and \citet{slater13} 
               (yellow squares for $>3\sigma$ detections and white squares for $<3\sigma$).
               White circles denote detections that can not be unambiguously tagged as 
               leading or trailing.}
       \label{fig:DistVsRA}
   \end{figure*}

   The resulting distances and distance uncertainties for these fields can be found in 
   table~\ref{table:distSgr}, 
   together with the central position of each field (in equatorial coordinates), a 
   'faint/bright branch' tag (derived from figure~\ref{fig:mapPointings}), a tentative 
   classification as leading or trailing arm where 
   posible (see below), a 'primary/secondary detection' tag and the distance modulus ($\mu$). 
   In figure~\ref{fig:DistVsRA} we compare our results to values from the 
   literature~\footnotemark[4], split in two diagrams (top panel for the faint branches and 
   bottom panel for the bright branches in both hemispheres). 
   \footnotetext[4]{The SDSS-DR8 measurements shown in this paper for the southern bright 
   arm have been corrected for the difference in the calibration of the red clump absolute 
   magnitude, as pointed out in \citet{slater13} and corrected in \citet{kopos13erratum}).
   And the SDSS-DR5 measurements have been decreased by $0.15 \ \mathrm{mag}$ to match the 
   BHB signal from SDSS, as prescribed in \citet{belok13}.}  

   Remarkably our turnoff point distances are not only in agreement with previous distance 
   measurements to known wraps, but also compatible with the distance predictions for nearby 
   wraps by the models of \citet{penarrub10} and \citet{law10}. In the following section 
   we discuss in detail these findings.
   
   \subsection{Comparison with models of the Sgr stream}\label{subsec:compareModels}

\begin{figure*}
   \centering
   \includegraphics[width=\textwidth]{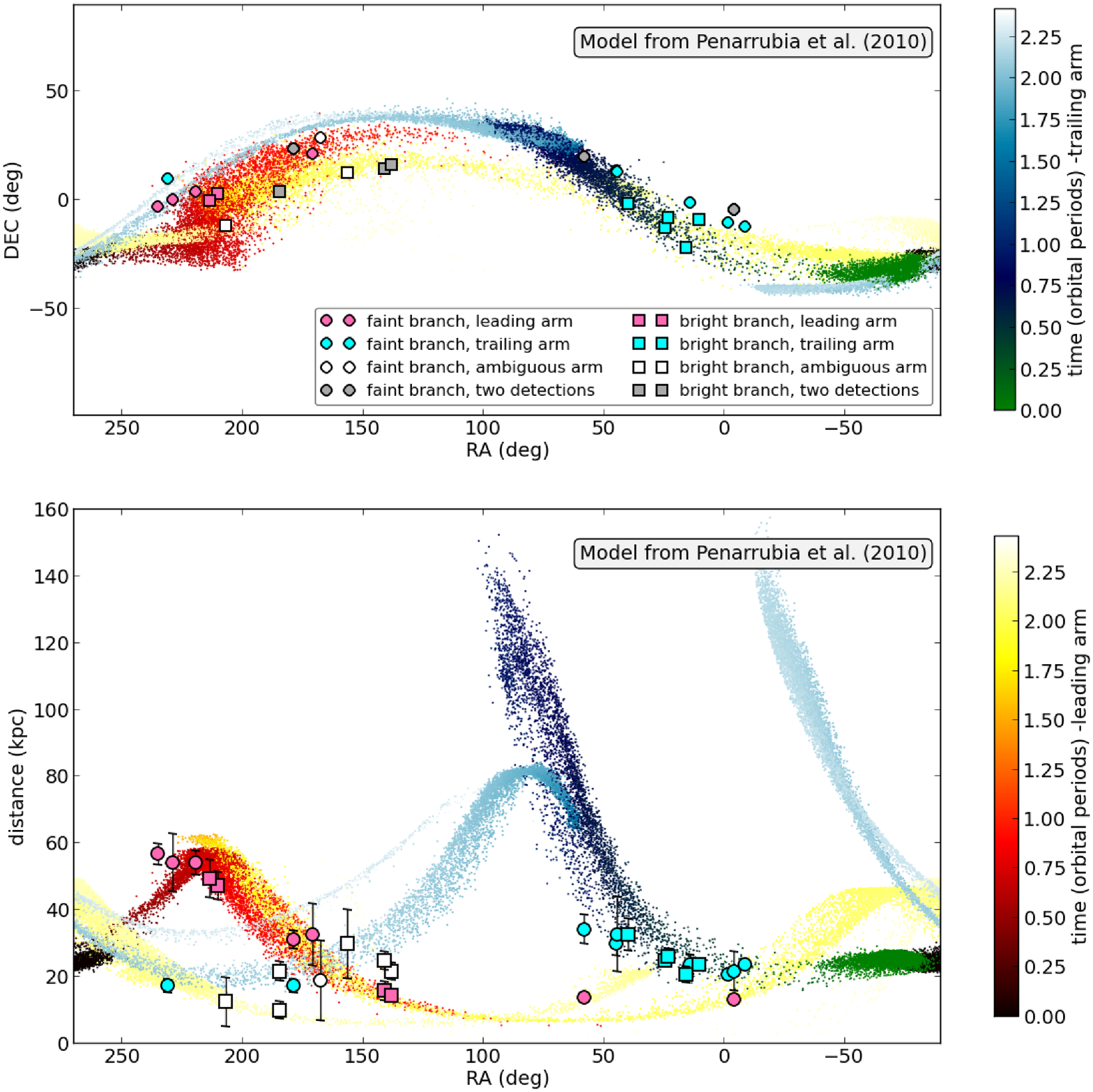}
      \caption{Our data compared to the predictions by the model from \citet{penarrub10}. 
               \emph{Top panel}: Equatorial map with the position of our fields plotted 
               over the simulation. 
               \emph{Bottom panel}: Distance vs RA diagram with our results compared to the 
               model predictions. 
               Fields on the faint branch are denoted with circles, and fields on the bright 
               branch are denoted with squares.
               Measurements matching the leading arm are denoted in pink, whereas those matching 
               the trailing arm are denoted in light blue. White markers represent detections 
               that can not be unambiguously tagged as leading or trailing; grey markers in the 
               upper panel correspond to fields with more than one MS detection (they unfold in 
               the bottom panel). 
               The colour scales represent the time since the particles from the 
               simulations became unbound.}
       \label{fig:modelJorge}
   \end{figure*}

\begin{figure*}
   \centering
   \includegraphics[width=\textwidth]{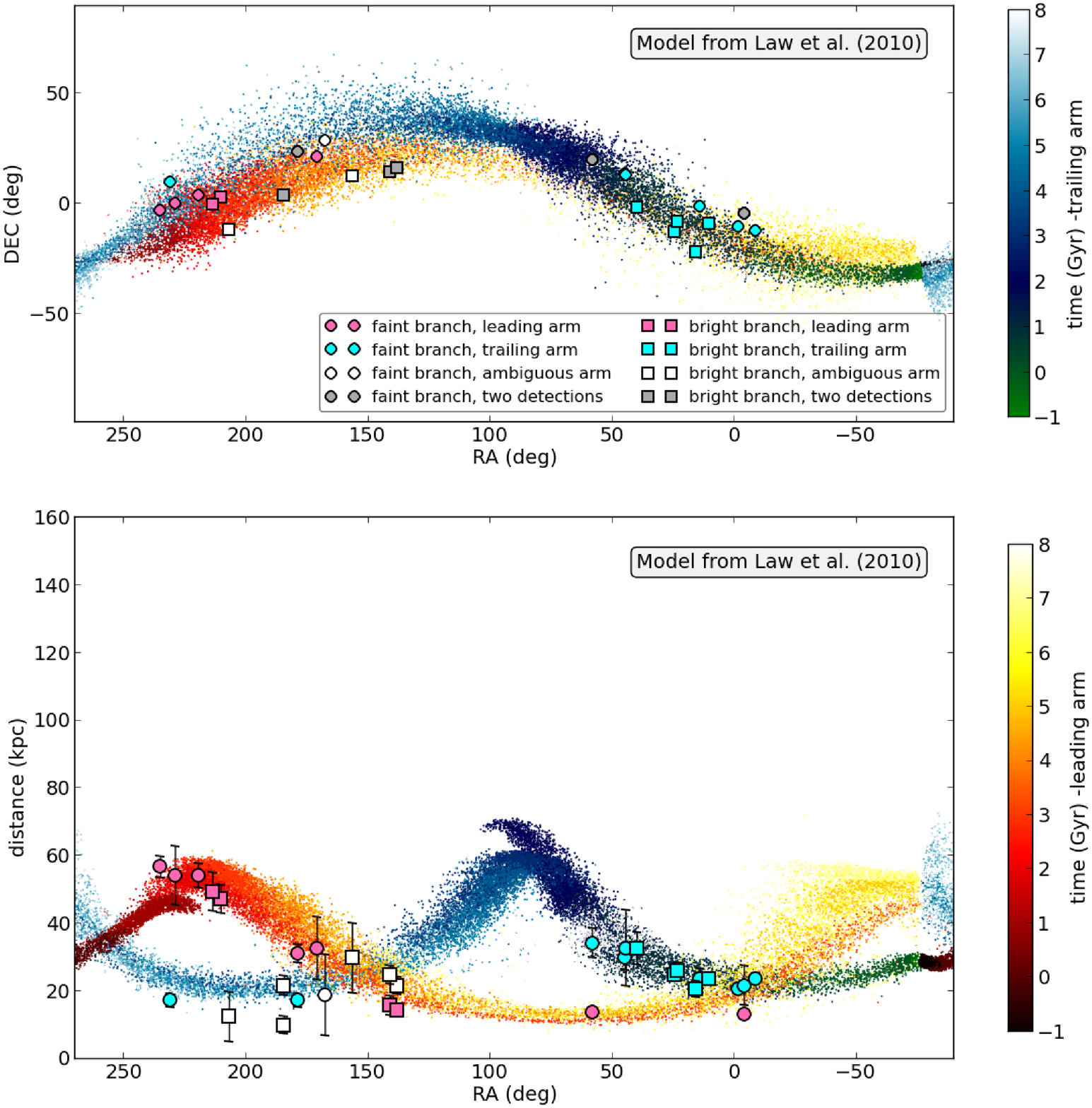}
      \caption{Same as in figure~\ref{fig:modelJorge} but for the model from \citet{law10}.}
       \label{fig:modelLaw}
   \end{figure*}

   Using the model predictions shown in figures~\ref{fig:modelJorge} and \ref{fig:modelLaw}, 
   we classify each field as belonging to the leading or trailing arm, by matching the distance 
   and the sky position.

   Of these two models, the model by \citet{penarrub10} seems to recover better the separation 
   in stellar density distribution that gives rise to the northern bifurcation into faint 
   and bright branches (figure~\ref{fig:modelJorge}, upper panels), whereas the model by 
   \citet{law10} seems to reproduce better the projected 2MASS stellar density distribution 
   (figure~\ref{fig:modelJorge}, lower panels). As noted in \citet{belok13}, the northern-
   trailing arm is more distant and has a steeper distance gradient than predicted by any 
   Sgr model. And although neither clearly recovers the southern bifurcation, they succeed in 
   reproducing the general distribution observed in that section of the stream. 
   
   \subsubsection{Northern leading arm}\label{subsec:SgrNorthLead}
   
   From our eighteen measurements on the bright and the faint branches of the northern-leading arm  
   (branches A and B, in the terminology of \citet{belok06}), nine clearly reproduce the 
   distance trends of \citet{belok06} and \citet{belok13} based on red giant and blue 
   horizontal branch stars (blue circles in figure~\ref{fig:DistVsRA}).
   For the faint branch, we extend westwards the distance measurements beyond those of 
   SDSS, and we provide its most distant detections so far --out to $56\mathrm{kpc}$ at
   RA~$\sim235^{\mathrm{o}}$. Comparing these most distant detections to the distance trend 
   of the models and to the bright branch at a similar right ascension, one can argue that these 
   detections likely lie close to the aphelion of the faint branch (or represent the aphelion 
   themselves), and therefore they are probably a good estimation for its distance.
   
   For the other nine detections, we find that the derived distances are either in mild disagreement 
   with the trends of the leading arm (four cases, white circles in figure~\ref{fig:DistVsRA}) or 
   incompatible with the leading arm (five cases, red and white circles in figure~\ref{fig:DistVsRA}). 
   In the single case of mild disagreement for the faint branch (A1185, RA~$\sim168^{\mathrm{o}}$) 
   the distance is well below the trends of both this and previous work (offset $\approx10 \ 
   \mathrm{kpc}$); however its large uncertainty prevents us from ruling out that 
   it belongs to the faint branch. We will discuss an alternative membership in subsection~
   \ref{subsec:SgrNorthTrail}. The three cases of mild disagreement for the bright branch 
   (ZwCl1023, A795-2 and A763-1, RA~$\sim150^{\mathrm{o}}$) are slightly above the distance 
   trend of this branch. Particularly, fields A795 and A763 also display two additional detections 
   (primary and secondary, respectively) slightly under the expected distance trend. Fields A795 and 
   A763 lie close in the sky (less than $4^{\mathrm{o}}$ apart) and both yield primary and secondary 
   distance measurements very consistent with each other and with this dichotomy. We interpret this
   as possibly indicating a region of the sky where the bright branch runs broader in distance.

   Out of the five detections incompatible with the distance trends of the leading arm, we will 
   discuss three (RXJ1524,A1413-2 and ZwCl1215-1) in 
   subsection~\ref{subsec:SgrNorthTrail}, together with the above mentioned A1185. Regarding the 
   other two (RXJ1347 and ZwCl1215-3, RA~$\sim205^{\mathrm{o}}$ and RA~$\sim185^{\mathrm{o}}$ 
   respectively), we have studied them individually and found the following. On the one hand, 
   ZwCl1215-3 matches the distance to the Virgo Overdensity \citep{bonaca12virgo} when using the 
   appropriate age and metallicity values for the theoretical isochrone, so it is likely a detection 
   of this cloud. On the other hand, RXJ1347 matches the distance and position predicted by the model 
   from \citet{penarrub10} for an older northern-wrap of the leading arm, but also the distances 
   predicted by the two models for the northern-trailing wrap. However we can not draw conclusions 
   regarding membership for an isolated detection and we lack kinematic data, so at the moment 
   we can not discriminate between both options (or even a third one).


   \subsubsection{Northern trailing arm}\label{subsec:SgrNorthTrail}

   In this subsection we revisit four detections in the galactic northern hemisphere which 
   yield distances incompatible with (or off) the leading arm. These detections are RXJ1524, A1413-2
   (red circles in figure~\ref{fig:DistVsRA}), A1185 (compatible with the faint 
   leading branch thanks to its large error bars, but severely offset from the distance trend)
   and ZwCl1215-1 (white marker at RA~$\sim185^{\mathrm{o}}$ on the bright branch). 
   The three detections in the faint branch (RXJ1524, A1413-2 and A1185) show distances strongly 
   consistent with each other ($\sim17 \ \mathrm{kpc}$). And the three of them are the fields 
   most apart from the Sgr orbital plane in our northern sample, spreading $60^{\mathrm{o}}$ along 
   the orbit.

   Remarkably both their distances and their positions in the sky are in extremely good agreement 
   with the predictions from the above mentioned models for the Sgr debris in the northern-trailing arm, 
   but at odds with the claim in \citet{belok13} that branch C (at lower declinations and more 
   distant) is indeed the continuation of the northern-trailing arm for this range of RA 
   (RA~$>160^{\mathrm{o}}$). In this sense it is worth noting that the model from \citet{penarrub10} 
   has predicted two nearly parallel branches for the northern-trailing arm (not only for the northern-
   leading arm). Therefore it is dynamically feasible that both the measurements for branch C 
   \citep{belok06} and the measurements in this work are tracing the trailing arm in the 
   galactic northern hemisphere, as far as they are probing two different branches.

   Given the consistency of our distance measurements with each other and with the simulations, and 
   given the distribution of the fields along the stream, we believe our detections in the faint branch 
   are a previously undetected part of a Sgr wrap, most likely a continuation of the section of the 
   northern-trailing arm presented in \citet{belok13}. However kinematic data or a spatially broader 
   photometric coverage are needed to confirm this. 

   Additionally, ZwCl1215-1, which lies on the bright branch, yields a distance measurement compatible 
   with the trend predicted for the northern-trailing arm. But its position on the sky (on the bright 
   branch) can not be reconciled with the current models for the trailing tail, neither with the 
   age, metallicity and distance values for the Virgo Overdensity. Thus, its membership 
   and meaning in the puzzle of the halo remain an open question.


   \subsubsection{Southern trailing arm}\label{subsec:SgrSouthTrail}

   Our measurements on the bright and the faint branches of the southern-trailing arm  
   reproduce the distance trends of \citet{kopos12, kopos13erratum} and \citet{slater13} 
   based on red clump and turnoff point stars. 
   For the faint branch, we confirm the trend set by the $<3\sigma$ detections in 
   \citet{slater13}, and we briefly extend westwards and eastwards the distance measurements. 
   Contrary to \citet{slater13}, we find no evidence for a difference in distance 
   between the faint and the bright branches of the southern-trailing tail. However it is 
   possible that such difference remains hidden in our distance uncertainties.
   
   When comparing to the above mentioned models, we find that the measures are in general 
   agreement with the predictions for both the faint and the bright branches. However the 
   distance gradient in the faint branch seems to be less steep in the data than in the 
   models, and the branch seems to be thinner in distance than predicted for any value of 
   the probed RA range. In this sense it is worth noting that, in contrast to what happens 
   to many of our northern hemisphere fields, only two of the CMDs in the southern 
   galactic hemisphere show secondary MS detections (RXJ0352 and RXJ2344, at 
   RA~$\sim58^{\mathrm{o}}$ and RA~$\sim356^{\mathrm{o}}$, respectively). And the 
   difference between the turnoff point brightness of these double detections does not 
   favour a thick branch, but rather the detection of a previously unknown nearby wrap (see 
   subsection~\ref{subsec:SgrSouthLead}).


   \subsubsection{Southern leading arm}\label{subsec:SgrSouthLead}

\begin{figure*}
   \centering
   \includegraphics[width=\textwidth]{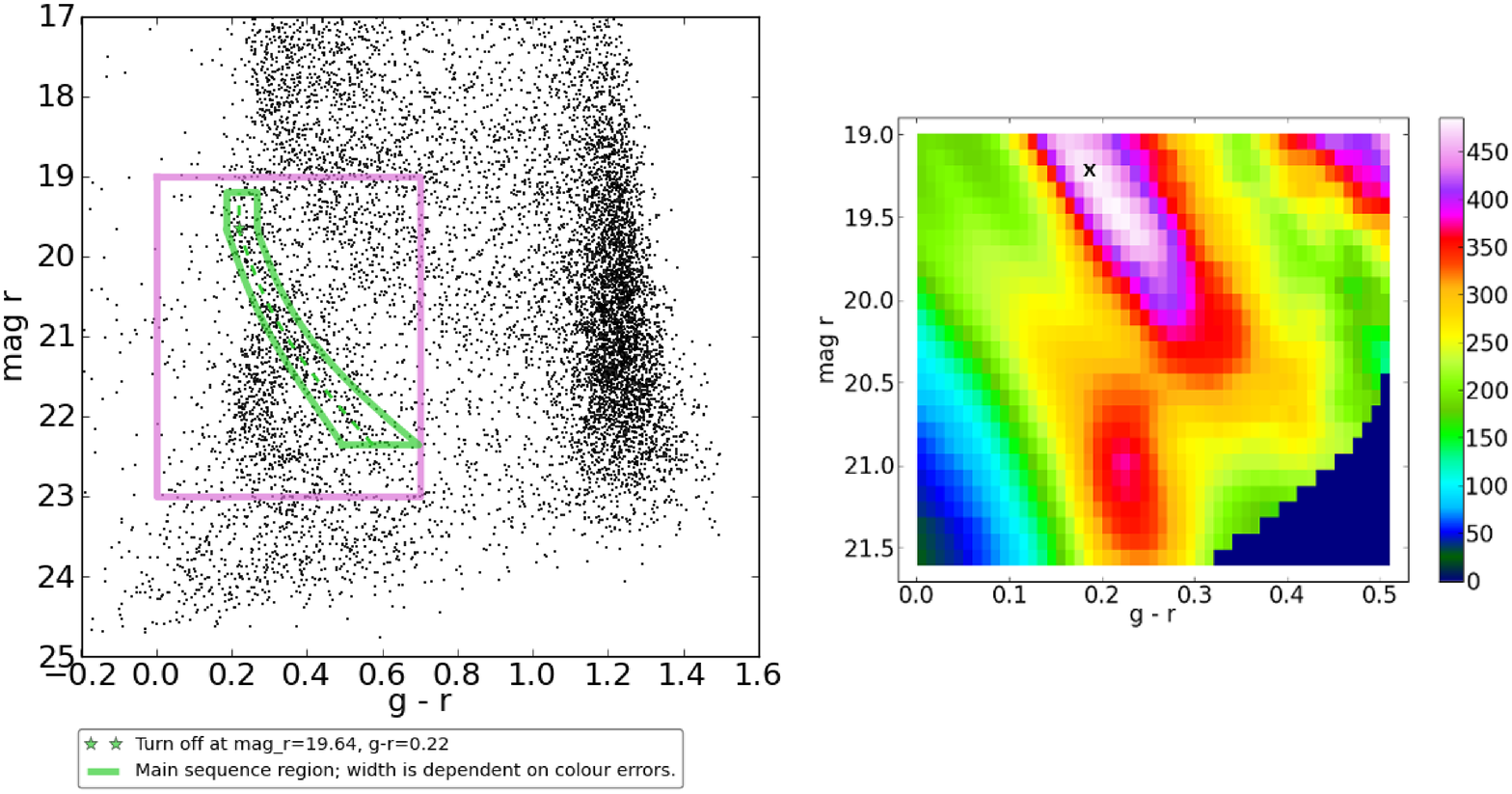}
      \caption{\emph{Left}: Dereddened CMD for the westernmost pointing probing the leading arm 
               in the southern hemisphere; the template main sequence function and the turnoff 
               point (green) are plotted for the maximum of the primary cross-correlation. 
               \emph{Right}: Weighted-density diagram resulting from the primary cross-
               correlation. The maximum (white bin, black cross) marks the top left corner of the 
               template-MS at the position of the southern-leading arm main sequence, whereas the 
               red overdensity at fainter magnitudes corresponds to the southern-trailing arm.
               }
       \label{fig:CMD-leadSouthRXJ0352}
   \end{figure*}

\begin{figure*}
   \centering
   \includegraphics[width=\textwidth]{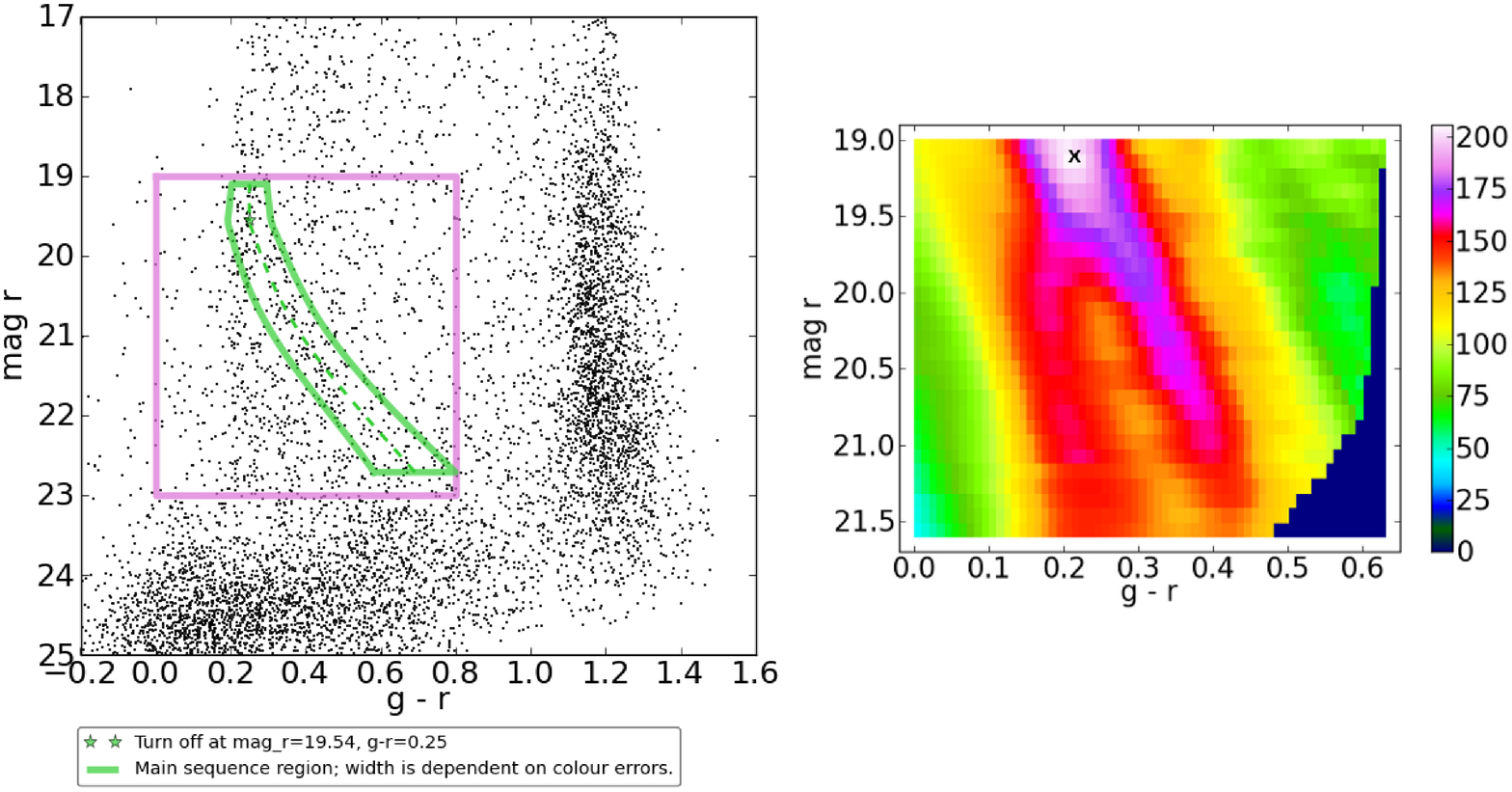}
      \caption{\emph{Left}: Dereddened CMD for the easternmost pointing probing the leading arm 
               in the southern hemisphere; the template main sequence function and the turnoff 
               point (green) are plotted for the maximum of the secondary cross-correlation. 
               We have randomly removed $50\%$ of the stars contributing to the primary 
               detection, which corresponds to the southern-trailing arm of the Sgr stream. 
               \emph{Right}: Weighted-density diagram resulting from the secondary cross-
               correlation. 
               The maximum (white bin, black cross) marks the top left corner of the template-MS 
               function at the position of the Orphan stream's main sequence. The primary 
               detection has been partially removed, and the remainings can be seen as a weak  
               tail at fainter magnitudes and slightly bluer colour.}
       \label{fig:CMD-leadSouthRXJ2344}
   \end{figure*}

   In this subsection we revisit the double detections of \ref{subsec:SgrSouthTrail}, namely 
   RXJ0352-1 and RXJ2344-2, (RA~$\sim58^{\mathrm{o}}$ and RA~$\sim356^{\mathrm{o}}$, primary 
   and secondary detections, respectively). We show their CMDs and their cross-correlation 
   density diagrams in figures~\ref{fig:CMD-leadSouthRXJ0352} and \ref{fig:CMD-leadSouthRXJ2344}. 
   We find that, using the same isochrone we have 
   used to derive distances to all the Sgr fields, both yield a distance of $\sim13 \ 
   \mathrm{kpc}$. These distances are in excellent agreement with the predictions from the two 
   simulations for the leading arm in the South and also with the trend set by the leading-northern data. 
   We thus claim to have detected the continuation of the northern-leading arm into the southern 
   hemisphere for the first time. The positions of these fields, however, suggest that the leading 
   arm dives into the southern hemisphere at higher declinations than predicted, overlapping in 
   projection with the faint branch of the trailing arm. 
   
   If the detection of the southern-leading arm or the northern-trailing arm proposed in this 
   paper are confirmed in future works (with kinematic measurements for membership or 
   photometric follow-up for spatial coverage), our measurements will be the closest and the oldest 
   debris of the Sgr stream detected to date. 
   If so, this would mean that our method has succeeded in detecting nearby substructure in 
   regions of the sky that had already been explored. The explanation to such a performance would 
   lie on the fact that we use a sample of stars (a large part of the main sequence) to identify 
   the overdensities in the CMD larger 
   than the sample of the usual halo tracers (red clump, red giants or blue horizontal branch), 
   and this could increase the contrast in regions of low concentration and thick disk contamination.

%
%

\section{The Palomar 5 stream and the Orphan stream}\label{sec:Pal5Orphan}


   \subsection{Turnoff point distances to the Pal5 stream and the Orphan stream}
   
\begin{figure*}
   \centering
   \includegraphics[width=\textwidth]{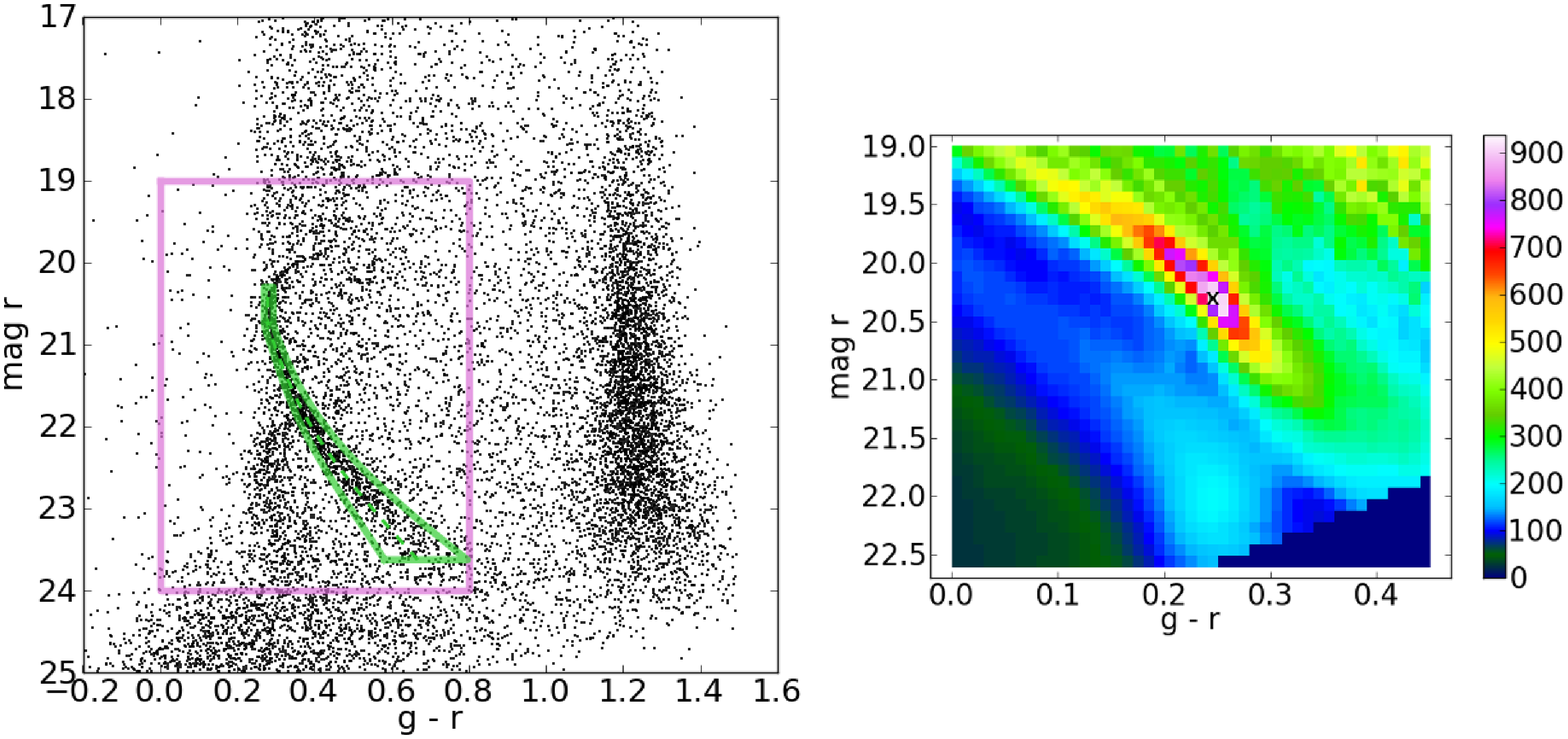}
      \caption{\emph{Left}: Dereddened CMD for the pointing containing the Palomar 5 stream as 
               its primary feature; the template main sequence function and the turnoff 
               point (green) are plotted for the maximum of the cross-correlation. The 
               secondary main sequence at fainter magnitudes corresponds to the faint arm 
               of the Sgr stream.
               \emph{Right}: Weighted-density diagram resulting from the cross-correlation. 
               The maximum (white bin, black cross) marks the top left corner of the template-MS 
               function at the position of the Palomar 5 stream's main sequence, whereas 
               the cyan overdensity at fainter magnitudes corresponds to the Sgr stream.}
       \label{fig:CMD-Pal5}
   \end{figure*}

\begin{figure*}
   \centering
   \includegraphics[width=\textwidth]{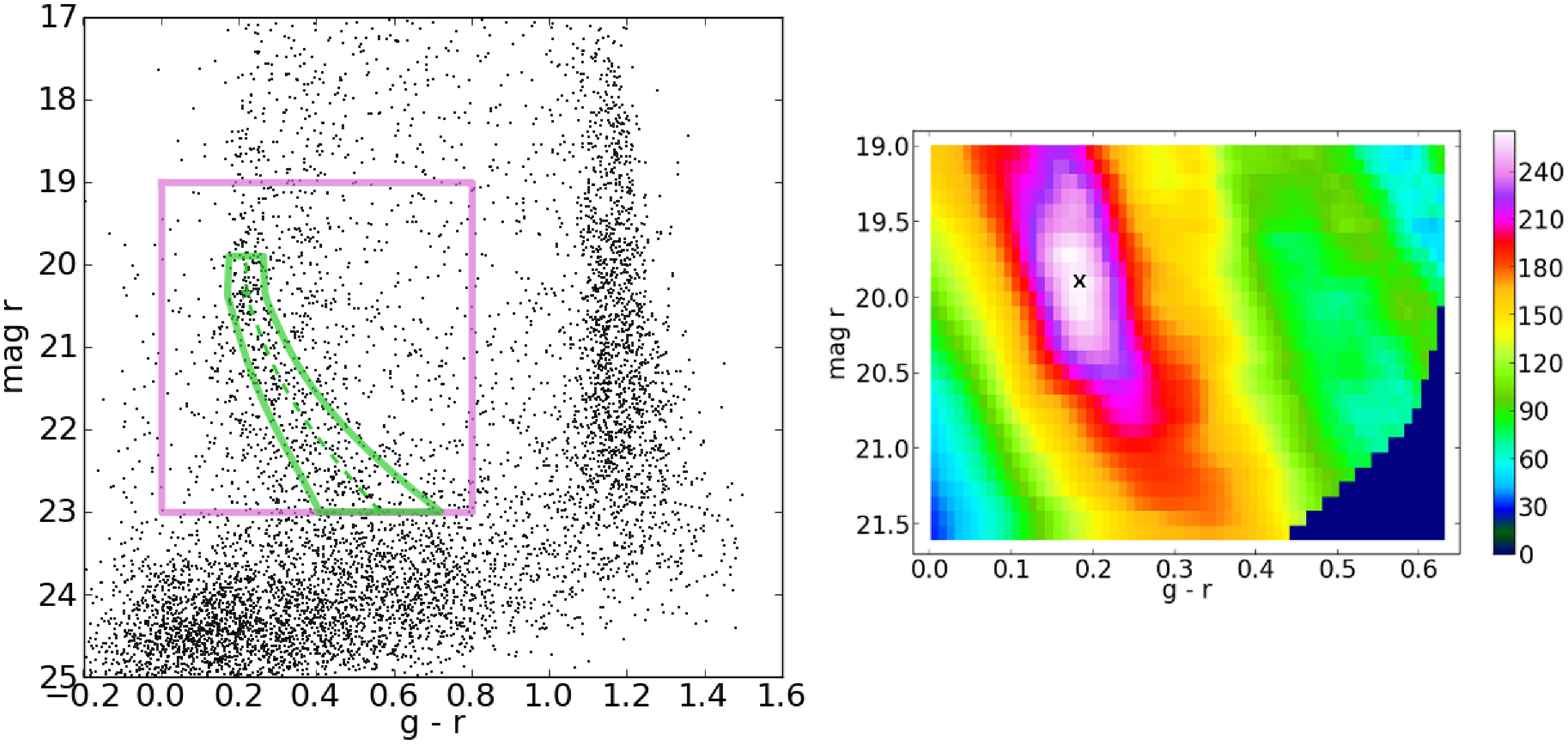}
      \caption{\emph{Left}: Dereddened CMD for the pointing containing the Orphan stream as 
               its secondary feature; the template main sequence function and the turnoff 
               point (green) are plotted for the maximum of the secondary cross-correlation. 
               We have randomly removed $60\%$ of the stars contributing to the primary 
               detection, which corresponds to the bright arm of the Sgr stream.
               \emph{Right}: Weighted-density diagram resulting from the secondary cross-
               correlation. 
               The maximum (white bin, black cross) marks the top left corner of the template-MS 
               function at the position of the Orphan stream's main sequence. The primary 
               detection has been removed, and thus it does not show in the density diagram.
               }
       \label{fig:CMD-Orph}
   \end{figure*}

   The Palomar5 stream and the Orphan stream are also probed by two of our fields (see 
   pink circles in figure~\ref{fig:mapPointings}). Their CMDs and their 
   corresponding turnoff points are shown in figures~\ref{fig:CMD-Pal5} and 
   \ref{fig:CMD-Orph}, together with their cross-correlation maps.

   We use these turnoff point values to calculate photometric distances to each of the 
   streams. Once again we assume single stellar populations characterized by theoretical 
   isochrones but now with $t_{age} = 11.5 \ \mathrm{Gyr}$ \citep{martell02} and 
   metallicity $[Fe/H]=-1.43$ \citep{harris96} in the case of the Pal5 stream, 
   and $t_{age} = 10.0 \ \mathrm{Gyr}$ and metallicity $[Fe/H]=-1.63$ in the case of 
   the Orphan stream \citep{casey13II}. These values correspond to measurements for 
   these particular streams, which are more metal-poor than the Sgr stream for similar 
   ages. Since the absolute brightness of the turnoff point for a given stellar population 
   depends on its age and metallicity, it is important to select representative 
   values in order to derive the right photometric distances.
   
\begin{figure*}
   \centering
   \includegraphics[width=\textwidth]{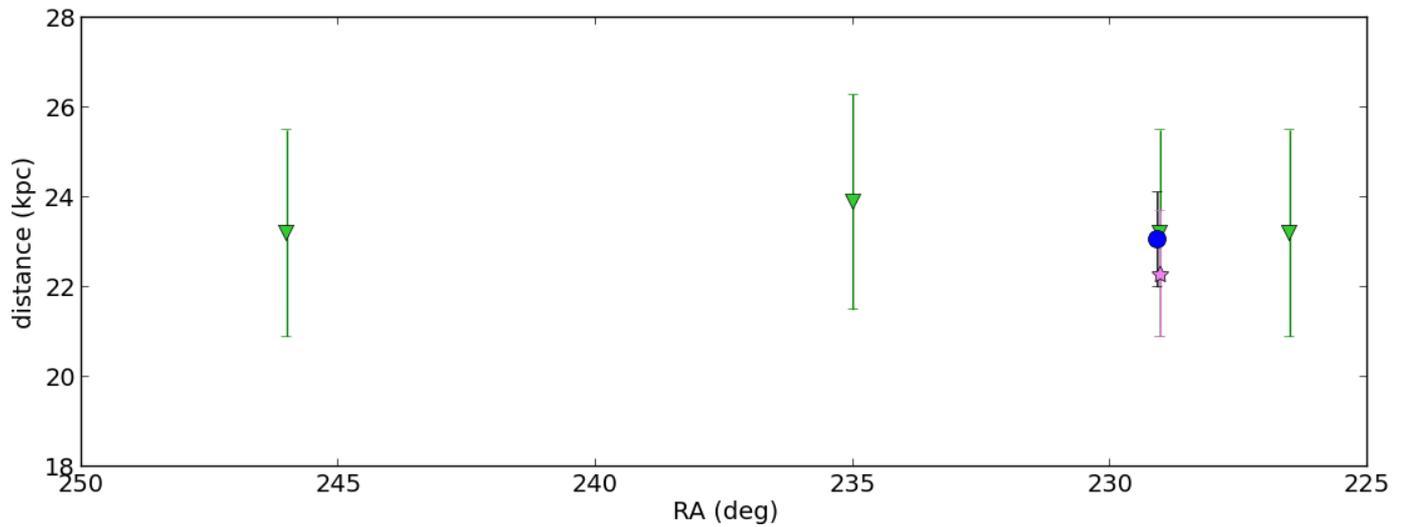}
      \caption{Photometric main sequence turnoff point distances along right ascension 
               for the Palomar5 stream. Our data point (blue circle) is based on a 
               single stellar population of age $11.5 \ \mathrm{Gyr}$ and metallicity 
               $[Fe/H]=-1.43$. The other values correspond to \citet{grilldion06pal5} 
               (green triangles) and \citet{vivas06} (pink star).
               }
       \label{fig:Pal5DistVsRA}
   \end{figure*}

\begin{figure*}
   \centering
   \includegraphics[width=\textwidth]{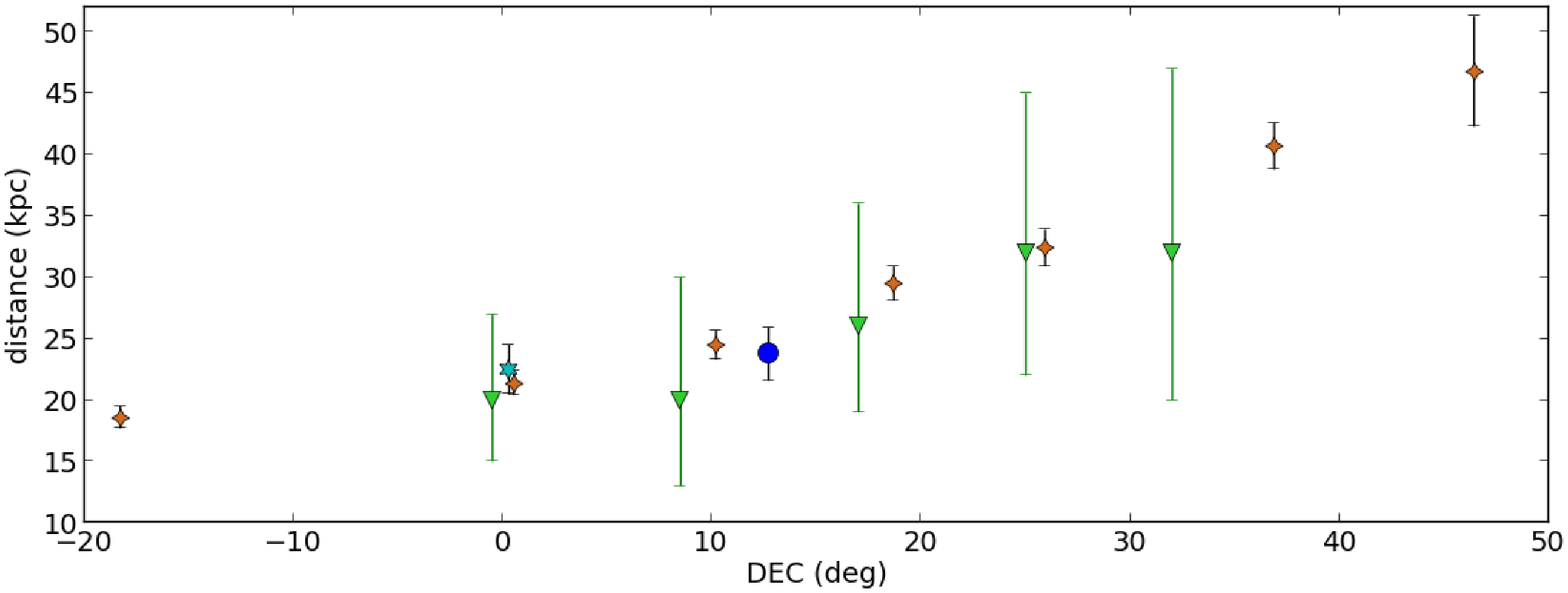}
      \caption{Photometric main sequence turnoff point distances along declination 
               for the Orphan stream. 
               Our data point (blue circle) is based on the theoretical isochrone  
               for a $10.0 \ \mathrm{Gyr}$ old stellar population with $[Fe/H]=-1.63$. 
               The other values correspond to \citet{belok07orphan} (green triangles), 
               \citet{newb10orphan} (orange diamonds) and \citet{casey13II} (cyan star).
               }
       \label{fig:OrphDistVsDEC}
   \end{figure*}

   The resulting distances are collected in table~\ref{table:distOthers} and displayed 
   in figures~\ref{fig:Pal5DistVsRA} and \ref{fig:OrphDistVsDEC}, respectively, where 
   they are compared to previous findings by other groups. Both results show good 
   agreement for the adopted age and metallicity values. 

   \begin{table*}
   \caption{Position and distances to the Palomar5 and Orphan streams:}    
   \label{table:distOthers}                       
   \centering                                                    
   \begin{tabular}{l c r c c c c}                            
   \hline\hline                                           
   Field & stream & RA (deg) & DEC (deg) & $\mu (mag)$ & $d$ (kpc) & $\Delta d$ (kpc) \\  
   \hline                                                  
      A2050 & Pal5 & 229.080749 & 0.08773 & 17.0 & 23.1 & 1.1 \\
      ZwCl1023 & Orphan & 235.040644 & -3.33158 & 16.6 & 23.8 & 2.2 \\ 
   \hline                                                              
   \end{tabular}
   \end{table*}

   \subsection{Influence of the age/Z isochrone values on the distances}\label{subsec:discussPal5Orph}

   For the Palomar 5 stream and the Orphan stream, our pencil-beam survey returns only 
   one detection each. We compare their derived distance measurements (table~
   \ref{table:distOthers}) to previous work (figures~\ref{fig:Pal5DistVsRA} and 
   \ref{fig:OrphDistVsDEC}, respectively) and find that our measurements are 
   consistent and well within the values to be expected from interpolations. We interpret 
   this as an independent validation of the stellar population parameters for these streams 
   in the literature: $11.5 \ \mathrm{Gyr}$ 
   and $[Fe/H]=-1.43$ for the Pal5 stream \citep{martell02,harris96}, and 
   $10.0 \ \mathrm{Gyr}$ and $[Fe/H]=-1.63$ for the Orphan stream \citep{casey13II}. 
   Variations in the absolute magnitude for the turnoff point of the 
   theoretical isochrone assigned to a given stellar population (characterized by a given 
   age and metallicity) propagate into the distance modulus, thus yielding variations in 
   the distance. For the Pal5 stream, our distance measurement can tolerate a relative 
   variation of $\Delta d_{rel}\approx0.15$ and still be in agreement with the previous 
   distance measurements; this variation threshold translates into an absolute magnitude 
   variation threshold of $\Delta M\approx0.35$. We find that variations in 
   $\Delta t=^{+1.7}_{-3.2}$ Gyr (limited by the formation time of the first stars) or 
   in $\Delta Z=^{+30}_{-6}\cdot10^{-4}$ dex (limited by the minimum metallicity available 
   in the set of theoretical isochrones) for the age and metallicity of the employed 
   theoretical isochrones meet this tolerance criterion. 
   For the Orphan stream, our distance measurement can tolerate a relative variation of 
   $\Delta d_{rel}=^{+0.24}_{-0.05}$, which translates into $\Delta M=^{+0.60}_{-0.11}$. 
   Variations in $\Delta t=^{+3.2}_{-1.3}$ Gyr or in $\Delta Z=^{+26}_{-3}\cdot10^{-4}$ 
   dex (same limitations as above) for the age and metallicity of the theoretical 
   isochrones respect this requirement.

%
%

\section{Conclusions}\label{conc}
   
   In this work we have used data from two deep cluster surveys, which provide randomly 
   scattered photometric pencil-beams in g' and r', and a field of view of $1\mathrm{deg}^2$ per 
   pointing. We have used this data to characterize previously 
   known substructure in the stellar halo of the Milky Way. 
   We analysed these data using two novel ingredients: a PSF-homogenization for 
   the images and a cross-correlation algorithm for the colour-magnitude diagram (CMD).
   The PSF-homogenization algorithm corrects the inhomogeneous distortion of the 
   sources across an image caused by the telescope's optics. In this 
   way, it recovers the true shapes and size distribution of the sources, improving 
   the performance of any star/galaxy separation procedure, specially at the 
   faint end. The cross-correlation algorithm explores the CMD of each 
   field searching for an overdensity with the shape of a stellar main sequence, 
   and returns the (colour,magnitude) coordinates of the corresponding turnoff 
   point, from which distances can be derived. Through this method, we have 
   shown that it is possible to exploit a two-filter pencil-beam survey to perform 
   such a study of streams or satellites, provided 
   that the contrast-to-noise ratio of the substructure's main sequence is moderately 
   significant. In this way our method bypasses the need for nearby control-fields 
   that can be used to subtract a reference foreground from the target CMDs.

   Using a set of theoretical isochrones \citep{marigo08,girardi10}, we have 
   calculated the distances to different regions 
   of the Sagittarius stream (faint and bright branches in both the northern and 
   southern arms) and obtained results in good agreement with previous work 
   \citep{belok06,kopos12,kopos13erratum,slater13} (see figure~\ref{fig:DistVsRA}). 
   
   We detect for the first time the continuation of the northern-leading arm into the 
   Southern hemisphere; we find that its distances are in excellent agreement with the 
   predictions by the models in \citet{penarrub10} and \citet{law10}, while the trajectory 
   seems to be located at higher declinations. 
   We also find evidence for a nearby branch of the northern-trailing arm at RA~$>160^{\mathrm{o}}$. 
   Both the distances and the footprint on the sky are in good agreement with the predictions 
   from the models. It is also compatible with being the continuation of the northern-trailing 
   region detected in \citet{belok13} if it turns or broadens to higher latitudes as it evolves 
   westwards, but it does not follow the same distance trend as branch C \citep{belok06}. However 
   it is feasible that both trends represent the trailing arm in the galactic northern hemisphere 
   if they belong to two different branches, as predicted in the model from \citet{penarrub10}.
   
   We have also used age and metallicity measurements from previous work \citep{martell02,
   harris96,casey13II}, to calculate distances to the Pal5 stream and the Orphan stream. 
   These distances are 
   in good agreement with the results in the literature \citep{grilldion06pal5,
   vivas06,belok07orphan,newb10orphan,casey13II}, attesting --together with the results 
   from the Sgr stream for previously known regions of the stream-- the robustness and 
   accuracy of the cross-correlation. 

   The methods presented in this paper open 
   the possibility of using deeper existing pencil-beam surveys (maybe originally 
   aimed for extragalactic studies) to measure accurate distances (or ages or 
   metallicities, provided that two of the three parameters are known) to streams, 
   globular clusters or dwarf galaxies. The existence of these pencil-beam surveys or 
   the reduced requirements of prospective ones, allow for more complete maps 
   of the Galactic halo substructure at reduced observational costs.

%
%

\begin{acknowledgements}
      
      B.P.D. iss supported by NOVA, the Dutch Research School of Astronomy.
      H.H. and R.vdB. acknowledge support from the Netherlands Organisation for 
      Scientific Research (NWO) grant number 639.042.814.
      
\end{acknowledgements}

%
%

\bibliographystyle{aa}  
\bibliography{references} 






   
  



%

\end{document}